\documentclass[preprint]{aastex}

\begin{document}

\title{The Detectability of Orphan Afterglows}

\author{Ehud Nakar, Tsvi Piran}
\affil{Racah Institute for Physics, The Hebrew University,
Jerusalem 91904, ISRAEL}
\author{Jonathan Granot}
\affil{Institute for Advanced Studies, Princeton, NJ 08540, USA}

\begin{abstract}

The realization that GRBs release a  constant amount of energy
implies that the post jet-break afterglow evolution would be
largely universal.  For a given redshift all afterglows should be
detected up to a fixed observer angle. We estimate the observed
magnitude and the implied detectability of orphan afterglows. We
show that for reasonable limiting magnitudes ($m_{lim}=25$) orphan
afterglows would  typically be detected from small ($\sim
10^\circ$) angles away from the GRB jet axis. A detected orphan
afterglow would generally correspond to a ``near-miss" of the GRB
whose jet was pointing just slightly away from us.  With our most
optimistic parameters we expect that 15 orphan afterglows will be
recorded  in the SDSS and 35 transients will be recorded in a
dedicated 2m class telescope operating full time for a year in an
orphan afterglow search. The rate is smaller by a factor of 15
for our ``canonical" parameters. We show that for a given
facility an optimal survey should be shallower, covering larger
area rather than deeper. The limiting magnitude should not be,
however, lower than  $\sim$ 23rd as in this case more transients
from on-axis GRBs will be discovered than orphan afterglows.
About 15\% of the transients could be discovered with a second
exposure of the same area provided that it  follows after 3, 4
and 8 days for $m_{lim}=23$, 25 and 27.

-

\end{abstract}

\section{Introduction}

The realization that GRBs are beamed with rather narrow opening
angles, while the following afterglow could be observed over a
wider angular range, led immediately to the search for orphan
afterglows: afterglows which are not associated with observed
prompt GRB emission. Rhoads (1997) was the first to suggest that
observations of orphan afterglows would enable us to estimate the
opening angles and the true rate of GRBs.  An expanding jet with
an opening angle $\theta_j$, behaves, as long as its Lorentz
factor $\gamma >\theta_j^{-1}$, as if it is a part of a spherical
shell (Piran, 1994). As it slows down and reaches $\gamma\approx
\theta_j^{-1}$ the jet quickly expands laterally (Rhoads, 1999)
producing a pronounced break in  its light curve. As time
progresses it can be observed over wider and wider observing
angles, $\theta_{obs}$. Dalal et al. (2002) have pointed out that
as the post jet-break afterglow light curves decay quickly, most
orphan afterglows will be  dim and hence undetectable. They also
suggest that the maximal observing angle, $\theta_{max}$, of an
orphan afterglow will be a constant factor times $\theta_j$.
Hence the ratio of observed orphan afterglows, $R_{orph}^{obs}$,
to that of GRBs, $R_{GRB}^{obs}$,  will not tell us much about
the opening angles of GRBs and the true  rate of GRBs,
$R_{GRB}^{true}$.

The observation that GRBs have a  constant amount of total energy
(Piran et al, 2001; Panaitescu and Kumar 2001; Frail et al.,
2001) and that the observed variability in the apparent
luminosity arises mostly from variation in the jet opening angles
leads to a remarkable result: The post jet-break afterglow light
curve is universal (Granot et al., 2002). We calculate this
universal post jet-break light curve, using both first principle
considerations and a calibration from the observed afterglows of
GRB990510 (Harrison et al., 1999; Staneck et al., 1999), and
GRB000926 (Harrison et al., 2001). Using this light curve we
estimate the maximal flux at an observing angle $\theta_{obs}$
from the jet axis. Using this flux we estimate the total number
of orphan afterglows that can be observed given a limiting
magnitude and the distribution of these orphan afterglows as a
function of $\theta_{obs}$ and the redshift, $z$.

The assumption that the total energy is constant implies that
orphan afterglows will be detected roughly up to a constant
observing angle $\theta_{max}$ (which is independent of
$\theta_j$, for  $\theta_j < \theta_{max}$). In this case
$R_{orph}^{obs} \propto R_{GRB}^{true}$ and can teach us about
the distribution of $\theta_j$ (Granot et al. 2002).

We describe our analytic model and the phenomenological fits to
the observations in \S 2. We present our estimates for the
observed rate of orphan afterglows  in \S 3. We stress that these
are idealized optimistic estimates that do not consider various
observational obstacles.  We do not consider how and whether
these transients could be identified as afterglows and
distinguished from other transients.  In \S 4 we compare our
estimates for the expected rate of orphan afterglows with the
capabilities of several surveys.

\section{The Model}
Consider an adiabatic  jet with a total energy $E$ and an initial
opening angle $\theta_j$.  We consider a simple hydrodynamic
model for the jet evolution (Rhoads 1999; Sari, Piran \& Halpern,
1999). Initially the jet propagates as if it were spherical with
an equivalent isotropic energy of $2E/\theta_j^2$:
\begin{equation}
E = (2\pi/3) \theta_j^2 R^3 \gamma^2 n m_p c^2 \ ,
\label{Spherical}
\end{equation}
where $n$ is the ambient  number density. The spherical phase
continues as long as $\gamma>\theta_j^{-1}$. At this stage the
jet expands sideways  relativistically in the local frame, with
$\gamma=\theta^{-1}$ and adiabaticity implies that
\begin{equation}
E = (2\pi/3)  R^3  n m_p c^2 \ , \label{jet}
\end{equation}
and the radius of the shock remains constant\footnote{More
detailed calculations show that $R$ increases slowly and $\gamma$
decreases exponentially with $R$ (Rhoads 1999; Piran 2000)}. Note
that the evolution during this phase is independent of
$\theta_j$, whose only role is to determine the time of the jet
break. If $E$ and $n$ do not vary from one burst to another then
the light curve during this phase will be universal, depending
only on the microscopic parameters (the equipartition parameters,
$\epsilon_{B,e}$, and the power law index $p$ of the electrons
distribution) of the specific afterglow. During both phases the
observed time is given by:
\begin{equation}
t=(1+z) R /4 c \gamma^2 \ . \label{time}
\end{equation}
Equations \ref{jet} and \ref{time} yield that the jet break
transition takes place at (Sari, Piran \& Halpern, 1999):
\begin{equation}
t_{jet}= 0.7(1+z) (E_{51}/n_0)^{1/3}(\theta_j/0.1)^2 \, {\rm days}\ .
\end{equation}
where $Q_x$ denotes the value of the quantity $Q$ in units of
$10^x$ times its (c.g.s) units. Due to relativistic beaming, an
observer  located at $\theta_{obs}$ outside the initial opening
angle of the jet ($\theta_{obs}> \theta_j$) will (practically)
observe the afterglow emission only at $t_\theta$ when $\gamma =
\theta_{obs}^{-1}$:
\begin{equation}
t_\theta=   A (\theta_{obs}/\theta_j)^2 t_{jet} \ . \label{ttheta}
\end{equation}
Roughly at $t_\theta$ the emission is also maximal. From
then on it decays in the same way as for an on-axis observer. The
factor $A$ in Eq. \ref{ttheta} is  of order unity. The value of A
is uncertain and it differs from one model to another. For
example, according to model 2 of Granot et. al (2002) $A<1$ (i.e.
the peak flux seen by observer at $\theta_{obs}$, is seen before
the jet opening angle is $\theta_{obs}$), while  $A>1$ according
to their model 1. Therefore, lacking a better knowledge  we use
$A=1$ throughout this paper.

The  synchrotron slow cooling light curve for the initial
(spherical) phase is given by Sari, Piran \& Narayan, (1998) and
modified by Granot \& Sari (2002). Sari Piran \& Halpern (1999)
provide temporal scalings for the maximal flux and the
synchrotron and cooling frequencies during the modified
hydrodynamic evolution after the jet break. Combining both
results [using the Granot \& Sari (2002) normalization for the
fluxes and typical frequencies] we obtain the universal post
jet-break light curve. For the optical band which is usually
above the typical synchrotron frequency, but can be either above
or below the cooling frequency, we find:
\begin{eqnarray}
F_{\nu>\nu_c}(t)= 460\left[g_0(p)/g_0(2.2)\right]10^{2.2-p\over
4}\epsilon_{e,-1}^{p-1} \epsilon_{B,-2}^{p-2\over 4}
 \nonumber \label{Fnu1a}\\
n_0^{-p-2\over 12} E_{50.7}^{p+2\over 3}
t_{days}^{-p}\nu_{14.7}^{-p/2}(1+z)^{p+2\over 2}D_{L28}^{-2}\
\mu{\rm Jy} \ , \\
F_{\nu<\nu_c}(t)= 170\left[g_1(p)/g_1(2.2)\right]10^{2.2-p\over
4}\epsilon_{e,-1}^{p-1} \epsilon_{B,-2}^{p+1\over 4}
 \nonumber \\
n_0^{3-p\over 12} E_{50.7}^{p+3\over 3}
t_{days}^{-p}\nu_{14.7}^{(1-p)/2}(1+z)^{p+3\over 2}D_{L28}^{-2}\
\mu{\rm Jy} \label{Fnu1b}\ ,
\end{eqnarray}
where  $D_{L}$ is the luminosity distance and $g_0(p)\equiv
10^{-0.56p}(p-0.98)\left[(p-2)/(p-1)\right]^{p-1}$, $g_1(p)\equiv
10^{-0.31p}(p-0.04)\left[(p-2)/(p-1)\right]^{p-1}$.  The cooling
frequency, for which equation \ref{Fnu1a} and \ref{Fnu1b} give
the same flux is:
\begin{eqnarray}
\nu_c = 3.62\times 10^{15}\left({p-0.98\over p-0.04}\right)^2
10^{(2.2-p)/2}\nonumber \\
\epsilon_{B,-2}^{-3/2} n_0^{-5/6}E_{50.7}^{-2/3}(1+z)^{-1}\ {\rm
Hz}\label{nu_c}\ .
\end{eqnarray}
Note that $\nu_c$ does not depends on $\theta_j$, and it remains
constant in time (after the break).

Using Eq. \ref{ttheta} we obtain the maximal flux that an
observer at $\theta_{obs}$ will detect:
\begin{eqnarray}
F_{\nu>\nu_c}^{max}(\theta_{obs}) = 1670\left[g_0(p)/g_0(2.2)\right]A^{-p}
\epsilon_{e,-1}^{p-1}\epsilon_{B,-2}^{p-2\over 4}
 \nonumber \\
n_0^{3p-2\over
12}E_{50.7}^{2/3}\nu_{14.7}^{-p/2}(1+z)^{1-p/2}D_{L28}^{-2}
\theta_{obs,-1}^{-2p} \ \mu{\rm Jy} \ ,\label{Fnuobs1a} \\
F_{\nu<\nu_c}^{max}(\theta_{obs}) = 620\left[g_1(p)/g_1(2.2)\right]A^{-p}
\epsilon_{e,-1}^{p-1}\epsilon_{B,-2}^{p+1\over 4}
 \nonumber \\
n_0^{p+1\over 4}E_{50.7}\nu_{14.7}^{(1-p)/2}(1+z)^{(3-p)/2}D_{L28}^{-2}
\theta_{obs,-1}^{-2p} \ \mu{\rm Jy}\label{Fnuobs1b}  \ .
\end{eqnarray}
One notices here a  very strong dependence on $\theta_{obs}$. The
peak flux drops quickly when the observer moves away from the
axis. Note also that this  maximal flux is independent of the
opening angle of the jet, $\theta_j$.

Once we know $F_\nu^{max}(\theta_{obs})$ we can estimate,
$\theta_{max}(z,m)$, the maximal observing angle at which an
afterglow is brighter than a limiting magnitude, $m$. We then
proceed and estimate the total number of orphan afterglows
brighter than $m$.

We can estimate the observed flux using the ``cannonical" values
of the parameters. However, these are uncertain. Alternatively,
we can use the observations of the afterglows of GRB990510
(Harrison et al., 1999; Stanek et al., 1999) and of GRB000926
(Harrision et al., 2001), both showing  clear jet breaks to
obtain an ``observational" calibration of $F_\nu(\theta_{obs})$.
 As the $\theta_{obs}$ dependence in Equations \ref{Fnuobs1a}
and \ref{Fnuobs1b} is similar we can write:
\begin{equation}
F(\theta_{obs}) = F_0 f(z) \theta_{obs}^{-2p} \ , \label{Fnumax2a}
\end{equation}
where $F_0$ is a constant and $f(z)=(1+z)^{1+\beta}D_{L28}^{-2}$
includes all the cosmological effects.   $\beta$ is the spectral
slope ($F_{\nu}\propto\nu^{\beta}$). The value of $\beta$ depends
on whether the observed frequency is below or above the cooling
frequency $\nu_c$: $\beta_{\nu<\nu_c}=(1-p)/2$ and
$\beta_{\nu>\nu_c}=-p/2$. For simplicity, we consider throughout
the paper only the case where $\nu>\nu_c$ and $\beta=-p/2$.

If all GRBs are similar with the same total energy, ambient
density and microscopic parameters ($\epsilon_{e,B}$ and $p$)
than $F_0$ is a ``universal" constant. However, its value is
uncertain. We need the observed flux and $\theta$ at a certain
time to determine $F_0$. We use the on-axis observed afterglows
at the break time, when an observer at $\theta_{obs} = \theta_j
A^{-1/2}$ (for a narrower jet) would observe a similar flux to an
on-axis observer (for the observed jet). Using the parameters for
GRB990510 and GRB000926 we estimate$F_0$ directly. For GRB990510
(Harrison et. al. 1999) we have $\theta_j=0.08$, $p=2.1$, $z=1.6$
and $m_{break}=20mag$,
 where $m_{break}$ is the R band magnitude of the afterglow at
the break time and $F_0(990510) = 0.01\,\mu Jy$. For GRB000926
(Harrison et. al. 2001)   $\theta_j=0.1$, $p=2.2$, $z=2$ and
$m_{break}=20mag$ and $F_0(000926)=0.03\, \mu Jy$. Both values are
rather close to the theoretical estimate with $E_{51}=0.5$, $n=1$,
$\epsilon_e=0.1$, $\epsilon_B=0.01$, $p=2.2$ which yields
$F_0=0.02\, \mu Jy$ and magnitude 20 at the break time (for a burst
at $z=2$ with $\theta_j=0.1$).

Granot et al. (2002) show, using a more refined simulation of the
off-axis light curve (their model II), that when the off-axis
light curve 'joins' the on axis light curve, the off-axis maximal
flux is a factor of few less than the on-axis flux. For example a
burst at $z=1$ with $E_{51}=0.5$, $n=1$, $\epsilon_e=0.1$,
$\epsilon_B=0.01$, $p=2.5$ is estimated in this model at 24th
magnitude at $\theta_{obs}=0.22$. This result corresponds to
$F_0=0.002\, \mu Jy$. According to the observations and the fact
that the off-axis maximal flux is a factor of few less than the
on-axis flux we use in the following a ``canonical" model of
$F_0=0.003\, \mu Jy$ and $p=2.2$. We should keep in mind that there
is a large uncertainty (factor of $\sim 10$) in the absolute value
of the flux. This uncertainty does not change $\theta_{max}$ that
 scales as $F_0^{1/2p}$ by a large factor. However, the overall
detection rate depends, almost linearly on $F_0$. The total
number of observed bursts depends not only on $\theta_{max}^2
\propto F_0^{1/p}$ but also on the duration, $t_{obs}$ (see Eq.
\ref{tobs} below)  which is $\propto \theta_{max}^2$, Together we
find that the rate is $\propto F_0^{2/p}$.

\begin{figure}{t}
{\par\centering \resizebox*{0.95\columnwidth}{!}
{\includegraphics{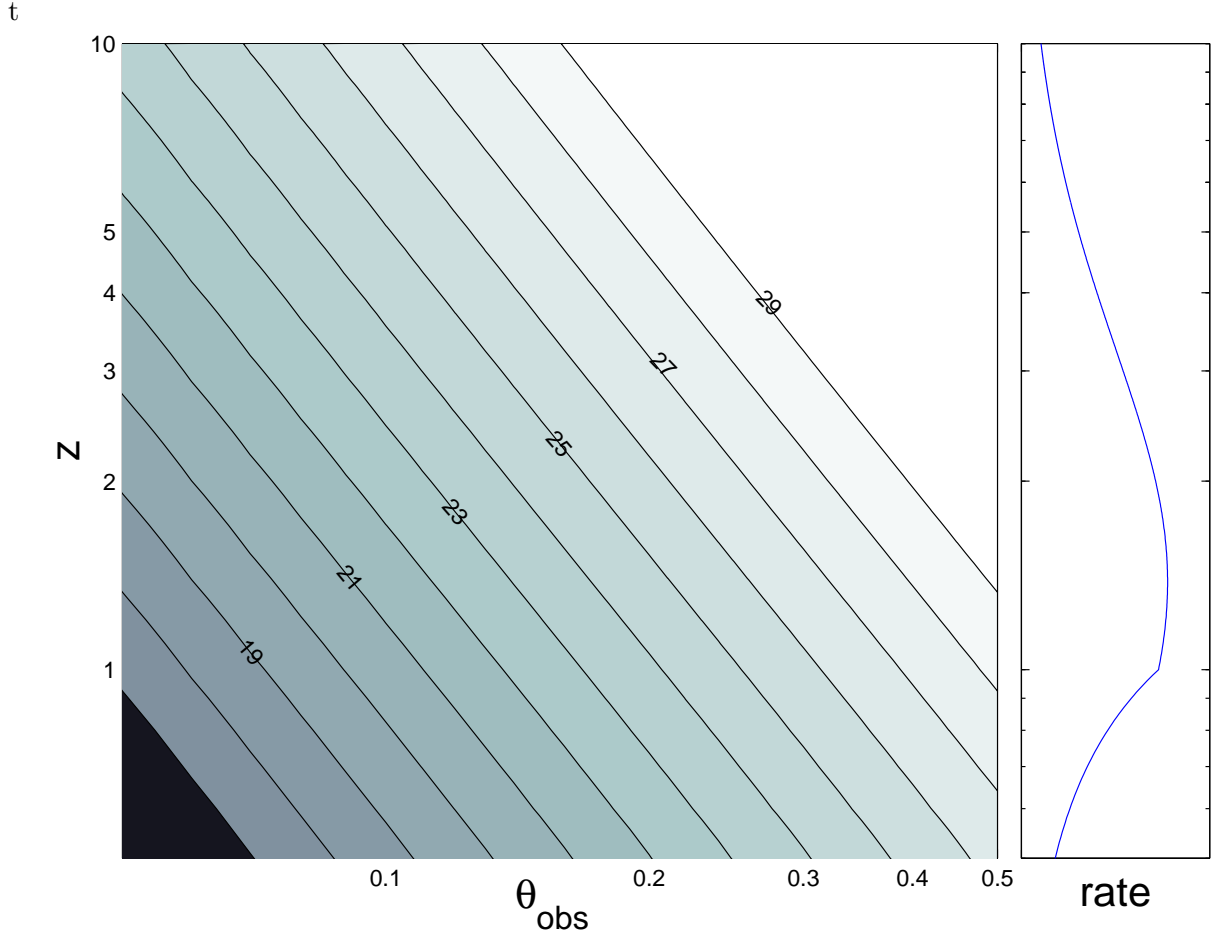}} \par} \caption{ \label{Contours} Left
panel: R magnitude contour lines (from 17- left bottom to 29) for
$F_0=0.003\,\mu{\rm Jy}$. The right panel depicts the GRB rate $
n(z) (dV/dz)/(1+z)$  as a function of $z$ (for $z_{peak}=1$), as
an indication to where we should expect most GRBs. As we plot only
the region with $z>0.5$ the flux depends almost as a power law on
the redshift.}
\end{figure}

For a given limiting magnitude, $m$, we can calculate now the
total number of observed orphan afterglows. For a given $z$ we
define $\theta_{max}(z,m)$  such that $23.6 -
2.5Log_{10}\{F_\nu[\theta_{max}(z,m)]\}=m$ where $F_\nu$ is the
observed flux in units of $\mu$Jy. Figure \ref{Contours} depicts
contour lines of the inverse function $m(\theta_{obs},z)$ on the
($\theta,z$) plane. We use throughout this paper a ``standard"
cosmological model with $\Omega_M=0.3$, $\Omega_\Lambda = 0.7$
and $h=0.7$.

According to our model an observer at $\theta_{obs}$ will observe
the orphan afterglow if $\theta_{obs}<\theta_{max}(z,m)$. The
afterglow will be detected over $t_{obs}(z,\theta,m)$:
\begin{equation}
t_{obs}(z,\theta,m) \approx  {A t_{jet} \over \theta_j^2}
(\theta_{max}^2 - \theta_{obs}^2)\label{tobs}
\end{equation}
until the signal decays below the limiting magnitude. In this
model which we denote model A, an orphan afterglow could be
observed from a solid angle\footnote{ The factor of 2 reflects
the expectation that the jets are double sided.} $2\pi
[\theta_{max}(z,m)^2-\theta_j^2]$  where $\theta_{max}(z,m)$ does
not depends on $\theta_j$. This model, and more specifically eq.
\ref{Fnumax2a}, are quite generally valid for
$\theta_{max}>\theta_j$.  However if $\theta_{obs}\sim\theta_j$,
$F_\nu^{max}(\theta_{obs})$  is more sensitive to the exact jet
model being used. Specifically, in such a case there will be a
significant contributions to the signal at $\theta_{obs}$ even
before the jet's opening angle reaches $\theta_{obs}$. This leads
to a larger flux at $\theta_{obs}$ and hence to a larger
$\theta_{max}$. For example, Granot et al. (2002) find, for their
 model II, that the solid angle between $\theta_j$ and the real
maximal angle ($\tilde \theta_{max}$) is approximately constant
even when $\theta_{obs}$ exceeds the asymptotic value of
$\theta_{max}$ (given by equation \ref{Fnumax2a}), i.e. $\tilde
\theta_{max}^2=\theta_{max}^2+\theta_j^2$. To take care of this
effect we consider also model B, in which we assume that this
relation holds. While model A is more conservative, model B
allows for more orphan afterglows. We see later that for
$m_{lim}\ge 25$ surveys, both models give comparable results.
Model B gives a factor of 2.5 more orphan afterglows for
$m_{lim}=23$ (with our "canonical" parameters). It also predicts
a larger ratio of orphan to on axis afterglows.

\section{Results}

The rate of observed orphan afterglows (over the entire sky) is
(in model A):
\begin{equation}
R_{orph}=\int_0^{10} {n(z)\over (1+z)} {dV(z)\over dz}dz
\int_{\theta_j}^{\theta_{max}(z,m)}  \theta d\theta  \ ,
\label{Rate1}
\end{equation}
where n(z) is the rate of GRBs per unit volume and unit proper
time  and  dV(z) is the differential volume element at  redshift
$z$.  We assume that the GRB rate is proportional to the SFR and
is given by:  \( n(z)=B~10^{0.75z} \) for \( z\leq z_{peak} \)
and \( n(z)=B~10^{0.75z_{peak}} \) for \( z_{peak}<z<10 \). The
normalization factor, $B$, is found by the condition: \(
R_{GRB}^{true}=f_{b}R_{GRB}^{obs}=\int _{0}^{10}(dV/dz)n(z)/(1+z)dz \)
where \( f_{b} \) is the beaming factor and $ R_{GRB}^{obs}=667\,
{\rm yr}^{-1}$ . In the following we consider two star formation
rates peaking at $z_{peak}=1$ and at $z_{peak}=2$ (Bouwenn, 2002).

Usually the detector's  exposure time is smaller than
$t_{obs}(z,\theta,m)$. Thus, the number of  detectable orphan
afterglows in a single snapshot,  over the whole sky, is:
\begin{eqnarray}
N_{orph}= \int_0^{10} {n(z)\over (1+z)} {dV(z) \over dz} dz
\quad\quad\quad\quad\nonumber\\
\times\int_{\theta_j}^{\theta_{max}(z,m)}   t_{obs}(z,\theta,m) \theta
d\theta \ . \label{Rate2}
\end{eqnarray}

If we consider now model B  we find:
\begin{eqnarray}
R^{(B)}_{orph}= {1\over 2}\left({F_0\over F_{lim}}\right)^{1/p}
\quad\quad\quad\quad\quad\quad\quad\nonumber\\
\times\int_0^{10}{n(z)\over (1+z)}{dV(z)\over dz}f(z)^{1/p} dz   \ ,
\label{Rate1_an}
\end{eqnarray}
where $F_{lim}$ is the limiting flux for detection. In this
case $R_{orph}/R_{GRB}^{true}\sim(F_0/F_{lim})^{1/p}$  is independent
of $f_{b}$. Using $t_{obs}(\theta_{obs})\approx
t_{\theta}(\tilde \theta_{max})-t_{\theta}(\theta_{obs})$, we estimate
the number of observed orphan afterglows:
\begin{eqnarray}
N^{(B)}_{orph}={A t_{jet}\over4 (1+z)\theta_j^2} \left({F_0\over
F_{lim}}\right)^{2/p}\quad\quad\quad\nonumber \\
\times\int_0^{10}n(z){dV(z)\over dz} f(z)^{2/p}dz \ . \label{Rate2_an}
\end{eqnarray}
We note that since $F_0\propto A^{-p}$ , both $R^{(B)}_{orph}$
and $N^{(B)}_{orph}$ scale as $1/A$.

\begin{figure}
{\par\centering \resizebox*{0.95\columnwidth}{0.3\textheight}
{\includegraphics{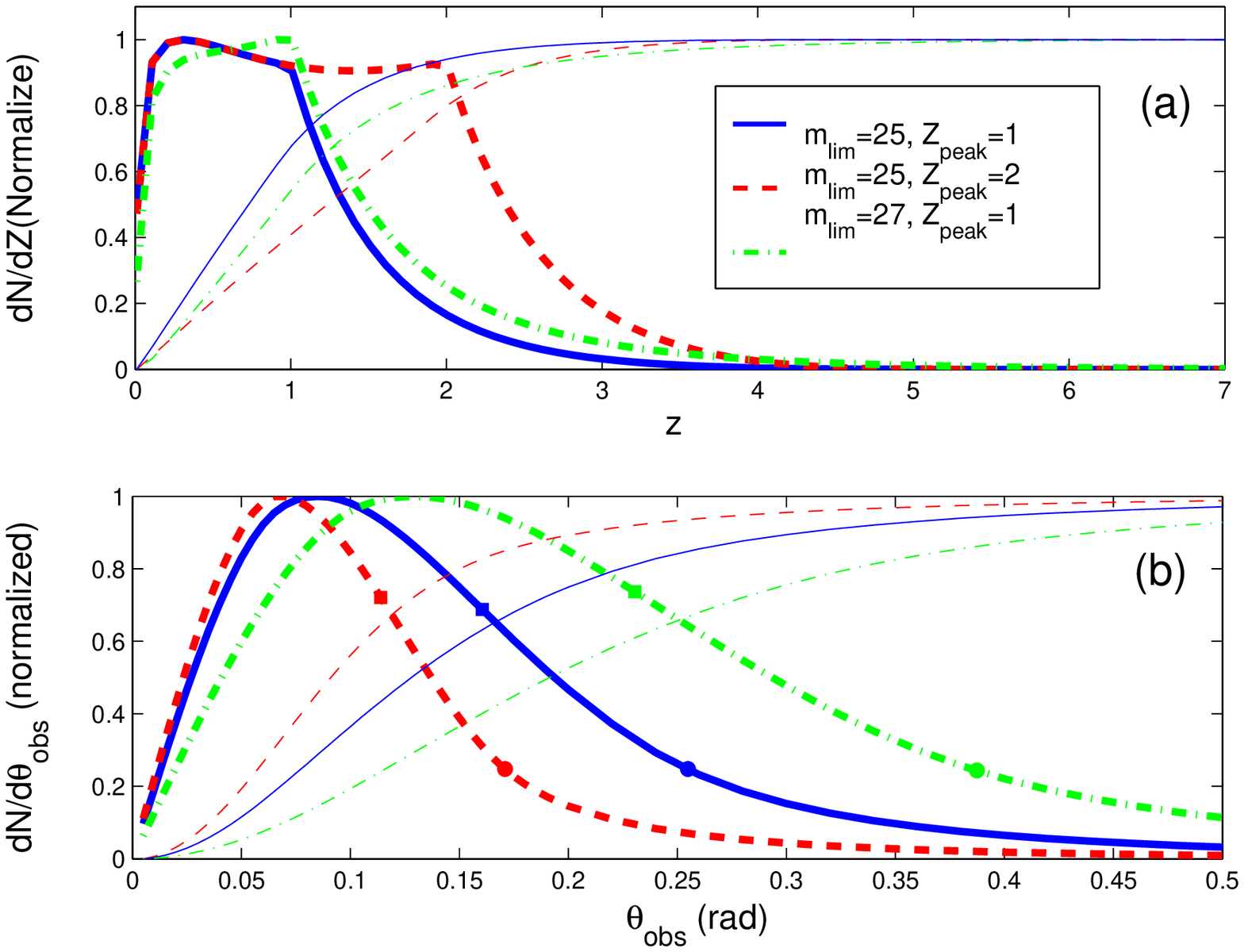}} \par}\caption{\label{dN-dz-dtheta} Top
panel (a): the normalized redshift  distribution of observed
orphan afterglows in a single snapshot ({\it thick lines}) and
the integrated $z$ distribution ({\it thin lines}). Bottom panel
(b): the angular distribution of observed orphan afterglows in a
single snapshot ({\it thick lines}) and the integrated
$\theta_{obs}$ distribution ({\it thin lines}). The circles
depict $\theta_{max}(z_{peak},m)$ and the squares depict $\langle
\theta_{obs} \rangle$. In both panels we use $F_0=0.003\,\mu$Jy,
and $\theta _{j}=0.1$. The different curves correspond to
$m_{lim}=25$, $z_{peak}=1 $ ({\it solid line}); $m_{lim}=25$,
$z_{peak}=2$ ({\it dashed line}); $m_{lim}=27$, $z_{peak}=1$
({\it dashed-dotted line}) and $m_{lim}=27$, $z_{peak}=2$ ({\it
dotted line}).}
\end{figure}

\begin{figure}
{\par\centering \resizebox*{0.95\columnwidth}{0.3\textheight}
{\includegraphics{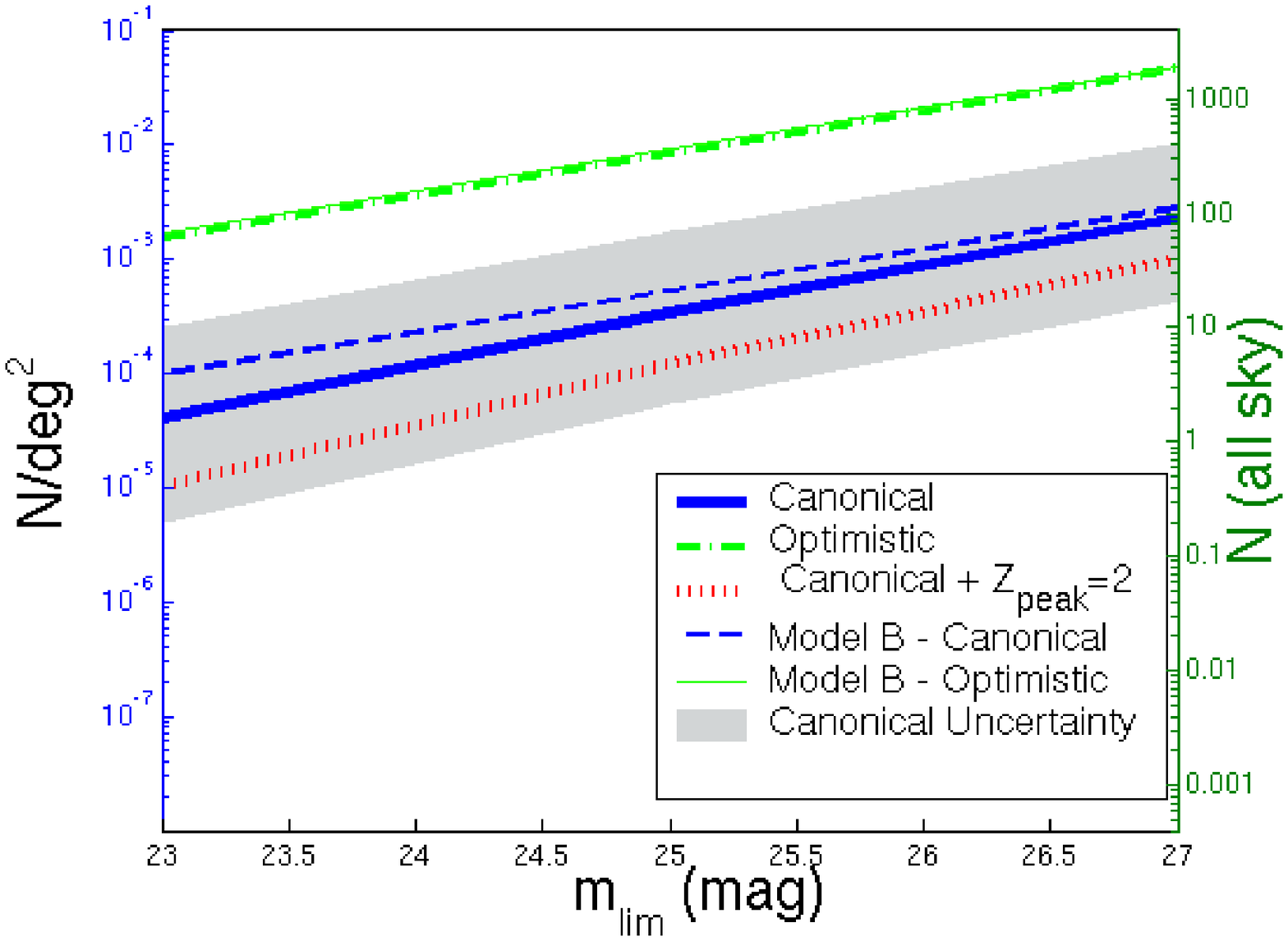}} \par} \caption{\label{Ntot} The number
of observed orphan afterglows per square degree (left vertical
scale) and in the entire sky (right vertical scale), in a single
exposure, as a function of the limiting magnitude for detection.
The  {\it thick lines} are for model A with three different  sets
of parameters: i) Our ``canonical" normalization
$F_0=0.003\,\mu{\rm Jy}$, $z_{peak}=1$, $\theta_j=0.1$ ({\it solid
line}). The gray area around this line corresponds an uncertainty
by a factor of 5 in this normalization. ii) Our most optimistic
model with a relatively small $\theta_j=0.05$ and a large
$F_0=0.015\,\mu{\rm Jy}$ ({\it dashed-dotted line}).   iii) The
same as our ``canonical" model, except for $z_{peak}=2$ ({\it
dotted line}). The {\it thin lines} are for model B, where the
{\it solid line} is for our ``optimistic" parameters, while the
{\it dashed line} is for our ``canonical" parameters. Both models
are similar for the ``optimistic" parameters while model B
predicts slightly more orphan afterglows then model A for the
``canonical" parameters.}
\end{figure}

The $z$ integrand of Eq. \ref{Rate2} gives the normalized redshift
distribution of the observed orphan afterglows for model A (see Figure
\ref{dN-dz-dtheta}a). This distribution has  a broad flat region all the
way from $z=0$ to $z_{peak}$. It sharply declines above
$z_{peak}$.  For a SFR with a higher $z_{peak}$ there are
significantly fewer orphan afterglows (see Figure \ref{Ntot}). The
function peaks (weakly) at $z_{peak}$ with higher limiting
magnitude as more orphan afterglows are observed in this case.
For model B the results are similar.

The $\theta$ integrand of Eq. \ref{Rate2}  [with
$\theta_{max}(z,m)$ replaced by $z_{max}(\theta,m)$ and
integration performed over $z$] yields the angular distribution
for model A (see Figure \ref{dN-dz-dtheta}b). For 25th limiting
magnitude the median observing angle is $5^\circ$-$7^\circ$
depending on the SFR rate. With lower $z_{peak}$ most GRBs are
nearer and hence stronger and detectable to a slightly larger
angles. These values are close to the jet opening angles. Most of
the observed orphan afterglows with this limiting magnitude will
be ``near misses" of on-axis events. The $\theta_{max}(z_{peak})$
are significantly larger ($10^\circ$ and $14^\circ$
respectively). With 27th magnitude the median of the angular
distribution moves way out to $12^\circ$. This is larger than
most GRB beaming angles but still narrow, corresponding only to
2\% of the sky. Again the results for model B are similar (as
long as $\theta_j \lesssim 0.1$).

As the sky coverage of GRB detectors is not continuously $4\pi$
it is possible that a GRB pointing towards us  (i.e. with
$\theta_{obs}<\theta_j)$ will be missed, but its on-axis
afterglow will be detected.  In Table I we provide the ratio of
the number of orphan afterglows to total number of afterglows
(both on-axis and orphan), in a single snapshot of the sky. As
expected this ratio is large for small jet opening angles and for
large limiting magnitudes  (and models A and B give similar
results in this limit) and it decreases with increasing
$\theta_j$ and decreasing $m_{lim}$.  An insensitive search
($m_{lim}=23$) yields for model A, a comparable (or even larger)
probability that a visual transient is a missed on-axis GRB whose
afterglow is detected or an orphan one. However, for model B, this
ratio is less sensitive to either $\theta_j$ or $m_{lim}$ for
$\theta_j\sim\theta_{max}$, and even for $m_{lim}=23$ the
majority of the afterglows would be orphans.
\begin{table*}
\begin{center}
\begin{tabular}{|c|c||c|c|c|}
 \hline    $z_{peak}$ &  $\theta_j$ & $m_{lim}=23$ & $m_{lim}=25$&    $m_{lim}=27$  \\
 \hline\hline
1 & 0.05 & 0.76 (0.81) & 0.88 (0.90) & 0.95 (0.95) \\ \hline 1 &
0.10 & 0.4\ \,\ (0.63) & 0.64 (0.73) & 0.81 (0.84) \\ \hline 1 &
0.15 & 0.2\ \,\ (0.55) & 0.4\ \,\ (0.63) & 0.64 (0.74) \\ \hline
2 & 0.10 & 0.2\ \,\ (0.56) & 0.44 (0.64) & 0.69 (0.76) \\ \hline

\end{tabular}
\caption  {Table I: The ratio of the number of orphan afterglows
to the total number of afterglows (both on-axis and  orphan), in a
single snapshot, for different limiting magnitudes and jet opening
angles.  The first value is for model A, while the second one in
brackets is for model B. }
\end{center}
\end{table*}

Figure \ref{Ntot} depicts the number of orphan afterglows per
square degree (and in the entire sky), in a single exposure, as a
function of the limiting magnitude. The thick lines are for model
A (with various parameters) while the thin lines are for model B.
In the ``optimistic'' case ($F_0=0.015\,\mu{\rm Jy}$,
$\theta_j=0.05$)  the predictions of both models are similar.
With the ``canonical" normalization ($F_0=0.003\,\mu{\rm Jy}$,
$z_{peak}=1$ and $\theta_j=0.1$) model B predicts 2.5 times more
orphan afterglows than model A,  for a limiting magnitude of 23.
These numbers provide  an upper limit to the rate in which orphan
afterglows will be recorded as point-like optical transients in
any exposure with a given limiting magnitude. These numbers
should be considered as upper limits, as our estimates do not
include effects such as dust extinction
that could  make these transients undetectable.
\begin{figure}
{\par\centering \resizebox*{0.95\columnwidth}{0.3\textheight}
{\includegraphics{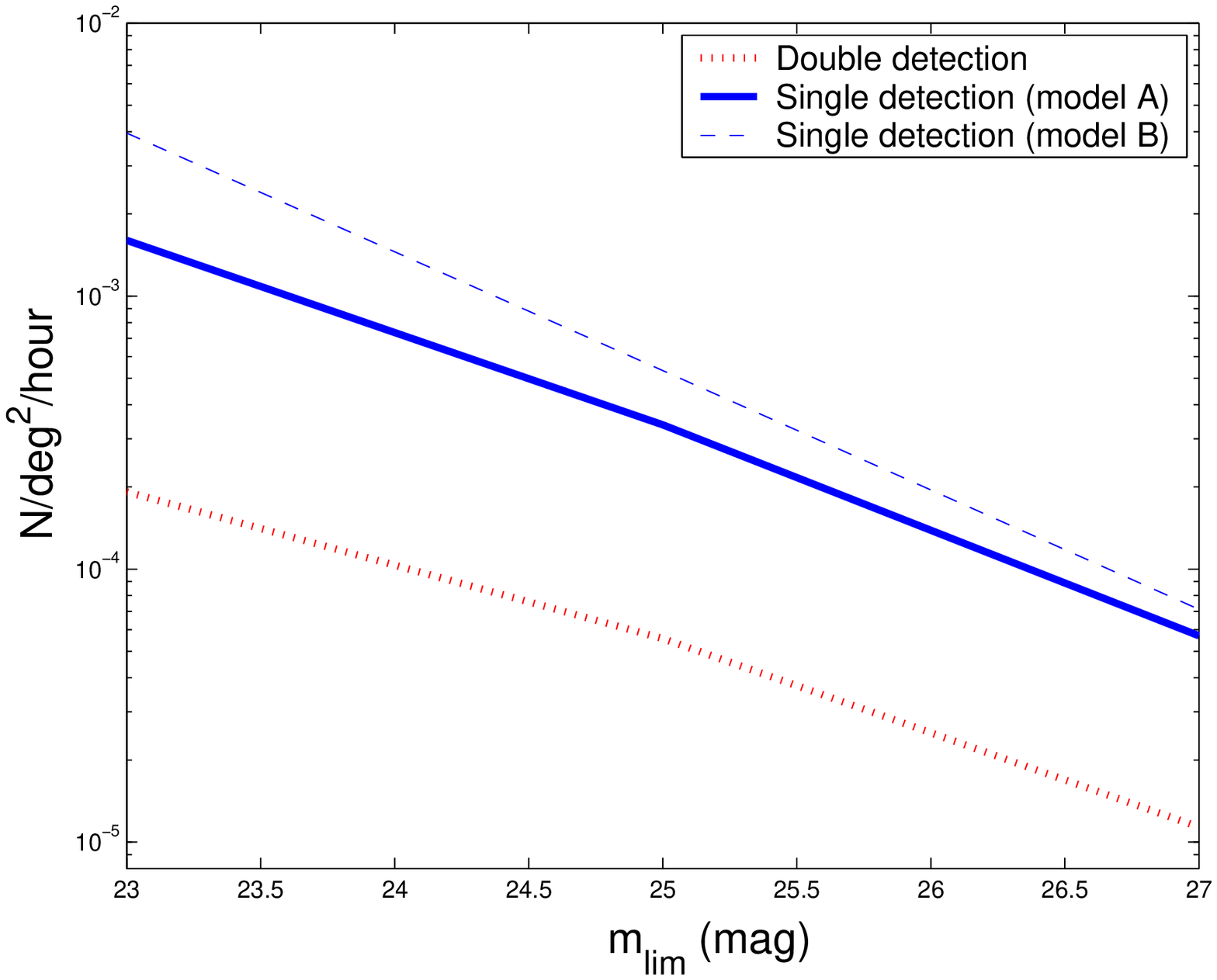}} \par} \caption{\label{DeepVsShal} The
detection rate of orphan afterglows per square degree per hour of
telescope time. The curves that correspond to the ``canonical"
normalization are for a single detection in model A ({\it solid
line}),  a single detection in model B ({\it dashed line}) and a
double detection in model A ({\it dotted line}). The rate is
calibrated for a telescope that reaches $m_{lim}=25$ in a one
hour exposure. It can be scaled up or down for other telescopes.}
\end{figure}

We  ask now  what will be the optimal strategy  for a given
observational facility: short and shallow exposures that cover a
larger solid angle or long and deep ones over a smaller area. The
 exposure time that is required in order to reach a given
limiting flux, $F_{lim}$, is proportional to $F_{lim}^{-2}$. We
can  now divide the number density of observed orphan afterglows
(shown in Figure \ref{Ntot}) by this time factor and obtain the
rate per square degree per hour of observational facility. Figure
\ref{DeepVsShal} depicts this rate with a calibration to a
hypothetical 2m class telescope that reaches a 25th magnitude
with one hour exposure. Other telescopes will have another
normalization but the trend seen in this figure remains. Figure
\ref{DeepVsShal} shows that shallow surveys that cover a large
area are preferred over deep ones (in both models). This result
can be understood as follows. Multiplying eq. \ref{Rate2_an} by
$F_{lim}^2$ shows that the rate per square degree per hour of
observational facility $\propto F_{lim}^{2-2/p}$.  As $p>1$
($2-2/p>0$) a shallow survey is preferred. A practical advantage
of this strategy is that it would be easier to carry out follow
up observations with a large telescope on a brighter transient
detected in a shallow survey. Additionally one can expect that
the number of spurious transients will be smaller in a shallower
survey. However, if one wishes to detect orphan afterglows, one
should still keep the surveys at a limiting magnitude of $\sim
23$, as for smaller magnitudes the number of  on-axis transients
detected will be comparable  to (model B) or even larger than
(model A) the number of orphan afterglows (see Table I).

\begin{figure}
{\par\centering \resizebox*{0.95\columnwidth}{0.3\textheight}
{\includegraphics{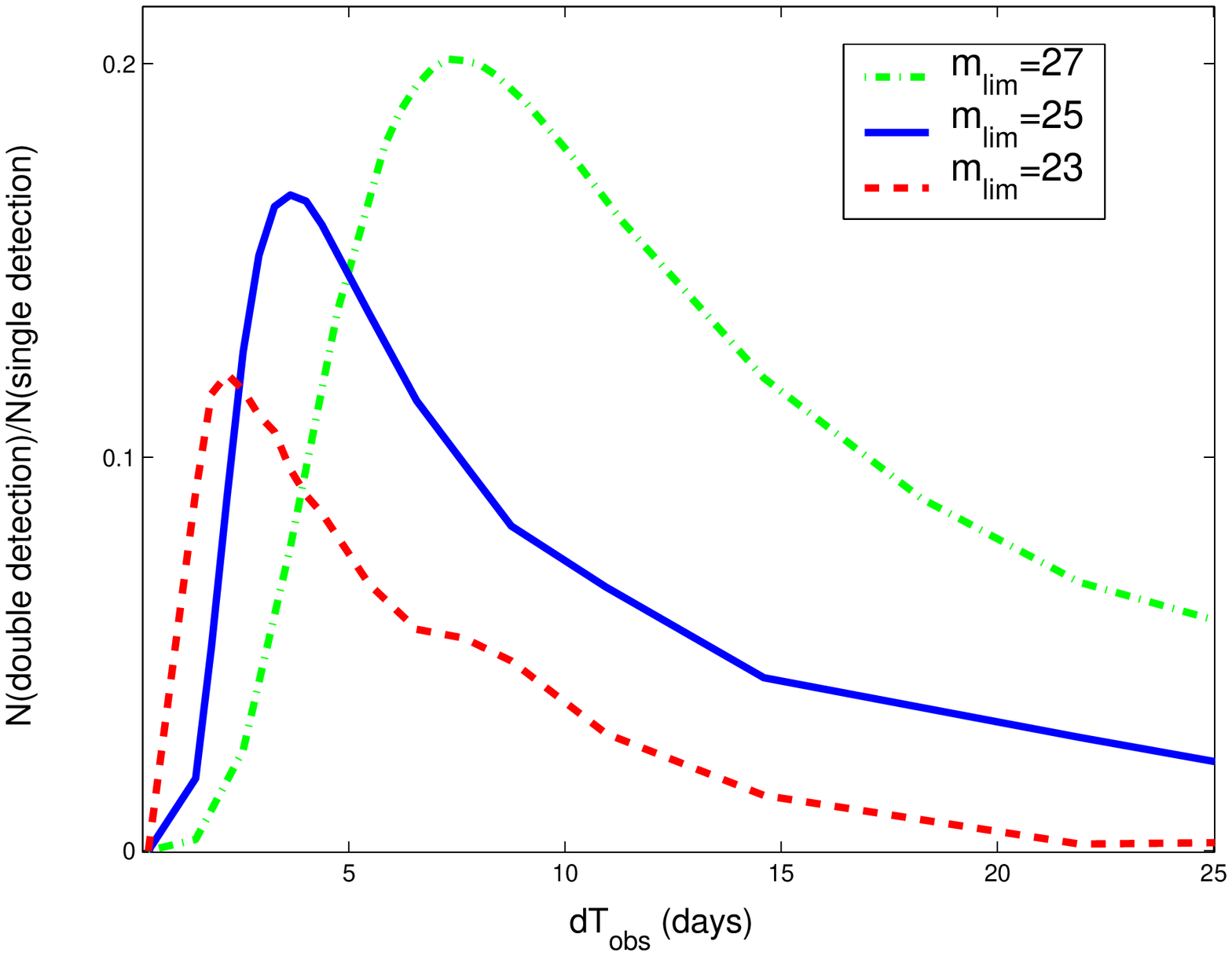}} \par}
 \caption{\label{Timedelay} The ratio
of double detection of  orphan afterglows to a single detection,
as a function of the time delay, $dT_{obs}$, between the two
exposures. The three models are: $F_0=0.003\,\mu{\rm Jy}$,
 $z_{peak}=1$
and $m_{lim}=23$ ({\it dashed line}) $m_{lim}=25$ ({\it solid line}) and
$m_{lim}=27$ ({\it dashed-doted line}).}
\end{figure}

The number density of orphan afterglow transients, given by Eq.
\ref{Rate2} and depicted in Figure \ref{Ntot} gives an indication
of the number of single  events recorded above a limiting
magnitude.  Clearly, a single detection is not enough in order
to identify that the transient indeed corresponded to an
afterglow. A second following detection of the transient with a
decrease by, say, 1 magnitude, would increase the probability 
that the transient is an orphan afterglow. We estimate the rate
of a double detection of this nature by an afterglow search with a
limiting magnitude $m_{lim}$ and a time delay $dT_{obs}$ between
the two exposures of the same region in the sky. We consider,
therefore, a detection of a transient in one exposure and a second
detection after time $dT_{obs}$ in which the  transient has
decayed  by more than one magnitude but is still above the
limiting magnitude of the survey. This is of course an idealized
situation and we do not consider realistic observational problems
(like weather etc.) which may make the detection rate lower. Not
surprisingly, the ratio of a double detection to a single
detection depends on the time delay, $dT_{obs}$ (see Figure
\ref{Timedelay}). The optimal time delay is the time in which an
afterglow from $z_{peak}$ and with $m_{lim}$ decays by one
magnitude: $dT_{obs}=3$ days for $m_{lim}=23$, $dT_{obs}=4$ days
for $m_{lim}=25$ and $dT_{obs}=8$ days for $m_{lim}=27$. With
this optimal choice, the fraction of the detected transients that
will be detected in a second exposure, is 20\% for $m_{lim}=27$
and 12\% for $m_{lim}=23$. Note however, that the curve becomes
narrower for low limiting magnitudes, making the exact timing of
the second exposure more critical. Deeper surveys are less
sensitive to the choice of the time delay.

It is interesting to compare the rate of observed orphan
afterglows to the true GRB rate as a function of the beaming
factor, $f_b \approx 2/\theta_j^2$.  At this point models A and B
differ the most. For  model B this ratio is $\propto
(F_0/F_{lim})^{1/p}$ and for the ``canonical" parameters it holds
the values of $0.004, 0.014, 0.031$ for $m_{lim}=23,25,27$
independently of  $f_b$.  In model A this ratio vanishes for low
$f_b$ values. For high $f_b$ values the ratio converges to
model's B constant ratio.  For the ``canonical" parameters this
ratio in model A reaches half of the asymptotic value  for
$f_b>50,125,250$ for $m_{lim}=27,25,23$ respectively. Thus, for
strong beaming (or high $m_{lim}$) it is possible to estimate the
true GRB rate from a determination of the orphan afterglow rate.
For  a low beaming factor this ratio is model dependent.

\section {Discussion}

We have calculated the number per unit solid angle of orphan
afterglows that could be detected by idealized surveys with
different limiting magnitudes. Our calculations are based on a
simple idealized model for the hydrodynamics of the sideways
expanding jets. Light curves from other models, including
numerical simulations (Granot et al., 2001) described in Granot
et al., (2002) show similar behavior. As the models differ  when
$\theta_{obs}\sim\theta_j$  we also consider an alternative
assumption for this range. This yields analytic expressions for
$R_{orph}$ and $N_{orph}$ (equations \ref{Rate1_an} and
\ref{Rate2_an}). We have shown that both models yield similar
results for surveys deeper than $m_{lim}=25$ and have minor
differences for surveys with $m_{lim} \sim 23$.

The normalization factor, $F_0$, of this light curve (Equation
\ref{Fnumax2a}) is somewhat sensitive to the choice of the model
parameters. Even observed afterglows with a clear jet break do
not yield a well defined normalization factor,
$F_0=A^{-p}f(z)^{-1}\theta_j^{2p}
F_{\nu}(t_{jet},\theta_{obs}=0)$, because of the uncertainties in
$\theta_j$, $A$, and $F_{\nu}(t_{jet},\theta_{obs}=0)$.  As
$N_{orph}\propto AF_0^{2/p}\propto
F_{\nu}(t_{jet},\theta_{obs}=0)^{2/p}/A$ (equation
\ref{Rate2_an}), we expect the uncertainty in $N_{orph}$ to be
similar to that in $F_0$ (a factor $\sim 10$). However, this
uncertainty does not influence the trends that indicate a
strategy for optimal orphan afterglow survey.

We stress that we do not consider here practical observational
issues, such as dust extinction or the ability to identify a weak
transient on top of a background host galaxy. We have also
assumed idealized weather conditions:  when considering a
``verified" identification by two successive detections we
assumed that a second exposure is always possible.  All these
factors will reduce the observed rate by some unknown factors
below our optimistic values. We do not consider how recorded
transients could be identified as afterglows. For example when we
discuss a single detection we mean a single photometric record of
a transient in a single snapshot. This is of course a very weak
requirement and much more (at least a second detection) would be
needed to identify  this event as an afterglow.

Finally one should be aware of the possible confusion between
orphan afterglows and other transients. Most transients could be
completely unrelated  (such as supernovae or AGNs) and could be
distinguished from afterglows with  followup observations.
However, there could be a class of related transients like
on-axis jets whose gamma-ray emission wasn't observed due to lack
of coverage or failed GRBs  (Huang et al. 2002). The only
quantitative measure that we provide is for the ratio of on-axis
afterglows to truly orphan ones  (i.e. off-axis afterglows whose
prompt gamma-ray emission would have been detectable if viewed
from within the jet opening angle).

We have shown that for a given dedicated facility the optimal
survey strategy will be to perform many shallow snapshots.
However, those should not be too shallow as we will  observe more
on-axis afterglows than off-axis ones. A reasonable limiting
magnitudes will be $\sim 23-25$. With this magnitude we should
perform a repeated snapshot of the same region after 3 to 4 days.

We consider now several possible surveys. The SDSS (York, 2000)
has $ m_{lim}\sim 23$ over $10^4$ square degrees in the north
galactic cap. Figure \ref{Ntot}  suggests that under the most
optimistic condition we expect 15 detection of  orphan afterglow
transients in the north galactic cap of the SDSS.  A careful
comparison of the SDSS sky coverage and the exposure of relevant
GRB satellites could reveal the rate of coincident detection (GRB
- Sloan optical transient) from which one could get a better
handle (using Table I) on the possible rate of detection of
orphan afterglows by SDSS. Under the more realistic assumptions
we expect a single orphan afterglow  transient in the SDSS.
Note that the 5 filters SDSS system could possibly select orphan
afterglow candidates even with only a single detection (Rhoads
2001; Vanden Berk et.al, 2001), though this method could be
applied only to bright sources. As the north galactic cap survey
provides only a single snapshot with no spectral information at
magnitude of 23,  even if a transient will be recorded in this
survey it will be impossible to identify it as a transient.
Clearly this is not the best survey for orphan afterglows search.
However, The south galactic cap survey, observes the same part of
the sky repeatedly with $m_{lim}\sim 23$. It can achieve a higher
magnitude, 25th, for steady sources. If the repetition interval
is longer than 2 weeks, than for most orphan afterglows, the
second observations can be considered as a new one. Therefore,
given that the time spent in the south cap is comparable to the
time spent in the north cup, the detection rate should be
comparable. The Advantage of the south cap is the ability to
identify transients sources even if the decay process could not
be observed,  and the possibility of ruling out sources that are
not transient in nature, such as AGNs.

Consider now  a dedicated 2m class telescope with an aperture $0.5
\Box^\circ$ of $m_{lim}=24$ for a 10 minutes exposure and
$m_{lim}=25$ for a 1 hour exposure. Under our most optimistic
assumptions it will record up to 35 orphan afterglows per year in
the shallow mode and a third of that (13 afterglows) in the deeper
mode. Using our ``canonical" model we find two orphan afterglows
per  year in the shallow mode.

The VIMOS camera at the VLT has a comparable aperture of $0.06
\Box^\circ$ but it can reach $m_{lim} =26$ in 10 minutes and
$m_{lim}=27$ in an hour. A single $m_{lim}=27$ frame would have a
$0.002$ orphan afterglows under the most optimistic assumptions
($1.1 \cdot 10^{-4}$ canonically). Thus, at best 2-3 orphan
afterglows should be hiding in every 1000 exposures. The
Subaru/Suprime-cum has a similar sensitivity but its field of view is
four times larger. Therefore every 1000 exposures should contain no
more than 10 orphan afterglows.It will be
however a remarkable challenge to pick up these transients and
confirm their nature from all other data gathered in these 1000
exposures.

We thank Tom Broadhurst, Re'em Sari and Avishay Gal-Yam for
helpful remarks. This research was supported by a US-Israel BSF
grant and by NSF grant PHY-0070928.

\end{document}